\documentclass{article}
\usepackage{fleqn,espcrc2}
\usepackage{epsfig,amsmath,color}

\DeclareMathOperator{\im}{Im}
\DeclareMathOperator{\re}{Re}
\DeclareMathOperator{\sign}{sign}
\begin{document}
\pagestyle{plain}
\title{Dispersive representation of $K\to 3\pi$ amplitudes and cusps\thanks{%
Presented by M.\ Z. and K.\ K. at Flavianet Kaon Workshop 08 and QCD 08. This work was supported in part by the Center for Particle
Physics (project no.\ LC 527), GACR (grant no.\ 202/07/P249) and by the EU
Contract No.\ MRTN-CT-2006-035482, \lq\lq{\sc Flavia\it net}''.
\newline In memoriam of our friend and colleague Jan Stern.
\newline $^\dagger$Unit\'e Mixte de Recherche  (UMR 6207)
 du CNRS  et des Universit\'es Aix--Marseille 1 et 2 et  Sud
Toulon--Var, Laboratoire affili\'e \`a la FRUMAM (FR 2291)
}
\begin{picture}(0,0)(0,0)
\put(7,55){\rm\small CPT-P170-2008}
\put(7,45){\rm\small PSI-PR-08-13}
\put(7,35){\rm\small UWThPh-2008-15}
\end{picture}
}

\author{K.~Kampf\address{Paul Scherrer Institut, Ch-5232 Villigen PSI, Switzerland}\address{IPNP, MFF, %
Charles University, V~Hole\v{s}ovi\v{c}k\'{a}ch 2, CZ-180 00 Prague 8, Czech Republic}, %
        M.~Knecht\address{Centre de Physique Th\'eorique$^\dagger$, CNRS-Luminy, Case 907, F-13288 Marseille Cedex 9, France}, %
        J.~Novotn\'y$^{\rm b}$, %
        M.~Zdr\'ahal$^{\rm b}$\address{Faculty of Physics, University of Vienna, Boltzmanngasse 5, A-1090 Vienna, Austria}, %
        }

\begin{abstract}
The NA48/2 collaboration has shown clear experimental evidence
for a cusp in the data for $K\to\pi\pi^0\pi^0$. This effect can be used
to extract information on the $\pi\pi$ scattering lengths. We address
this issue using a two-loop dispersive construction of $\pi\pi\to\pi\pi$
and $K\to\pi\pi\pi$ amplitudes in the presence of isospin breaking.
\end{abstract}

\maketitle


\section{Introduction}

The observation of a cusp anomaly in the $\pi^0\pi^0$ invariant mass
distribution in the data on the $K^\pm\to\pi^\pm\pi^0\pi^0$
decay collected by the NA48/2 collaboration \cite{na48} has triggered some
theoretical activity \cite{Cabibbo04}-\cite{bern}.
The basic explanation of the appearance
of a unitarity cusp is very simple \cite{Cabibbo04}: the amplitude for $K^{+}\to \pi ^{+}\pi ^0\pi
^0$ has two basic contributions, one of which corresponds to
the $\pi^{+}\pi^{-}$ intermediate state rescattering to $\pi ^0\pi ^0$.
This intermediate state in the s-channel generates (at the one-loop level) a square root singularity
\cite{meissner}, and the corresponding amplitude behaves at this level for $s\sim4m_{+}^2$ ($m_{+}$ is the mass of $\pi^+$) as
\begin{equation}
A(s)=R(\sigma_+ ^2)+\pi S(\sigma_+ ^2)\left\{
\begin{array}{l}
\!\!i\sigma_+ ,\ s>4m_{+}^2\\
\!\!\!-\widetilde\sigma_+,\ s<4m_{+}^2
\end{array}\!\!,
\right.   \label{cusp}
\end{equation}
where $
\sigma_P = \sqrt{1- 4m_{P}^2/s}$, ${\widetilde\sigma}_P =\sqrt{4m_{P}^2/s - 1}$. The functions  $R(\sigma_+ ^2)$ and $S(\sigma_+ ^2)$ can be expressed as
convergent series in $s-4m_{+}^2$ in the physical region of the $%
K^{+}\to \pi ^{+}\pi ^0\pi ^0$ decay. This singularity appears at $4m_{+}^2$,
which is above the physical threshold $4m_0^2$ ($m_0$ is the mass of $\pi^0$), and the cusp is a result of the interference of the part containing
the singularity and the rest without it. It is clear that the cusp
appears only in the isospin breaking case and its strength is sensitive to
the $\pi\pi $ scattering amplitude at the threshold.
This is the reason why the investigation of the cusp effect can in principle
serve as an independent method for the experimental determination
of the $\pi\pi $ scattering lengths, provided a model independent description
of the corresponding amplitude can be given.

In \cite{CI} Cabibbo and Isidori
have proposed to use the assumed simple analytical properties of the amplitude
($\pi\pi$ amplitudes are considered in the form (\ref{cusp}) with $R(\sigma_+^2)$ and $S(\sigma_+^2)$ polynomial)
and the unitarity of the scattering matrix in order to express the amplitude near
the threshold as an expansion in scattering lengths $a_i$.
This idea was further investigated in \cite{gamiz}.
Another possibility how to address this issue is the framework
of non-relativistic effective field theory as developed  in \cite{bern}.
This is done as a combined expansion again in $a_i$, and also in the pion momenta.
Both of these different
approaches compute contributions to the $K^{+}$ decay amplitude
up to order $O(a_i^2)$. Each of these approaches has its own limitations
that result from the assumptions made. Moreover, they give no
connection between cusp effects and the traditional PDG parameterization of the $K$
decay amplitudes. We consider an alternative approach, which rests on general properties,
unitarity, analyticity, crossing symmetry, relativistic invariance, and chiral power counting
for partial wave amplitudes. This leads, through a two-step iterative procedure,
to a two-loop representation of the $K\to\pi\pi\pi$ amplitudes.

In the following we illustrate the method on the simplest case of
the $K_L\rightarrow 3\pi ^0$ decay, mainly because
the analytical expressions are less involved, but also
because, for the time being, we want to avoid addressing some further issues appearing
in the treatment of the processes involving charged pions. The cusp in this
decay has been observed by KTeV \cite{neutral cusp KTeV}, as well as by NA48/2 \cite{na48-K_L}.

\section{Reconstruction theorem}

The approach we wish to implement proceeds in parallel with the construction
of the two-loop representation for the $\pi\pi$ scattering amplitude achieved
in \cite{stern,KMSF1}. We shall use the fact that it can be
extended to other processes and that isospin symmetry is not an
essential ingredient \cite{zn}. Here we are interested in
$K\pi\to\pi\pi$, related to $K\to 3\pi $ by crossing symmetry. The essential ingredients required
in order to implement this construction are the following.
First, we need a decomposition of the amplitude of the type
\begin{equation}
\mathcal{A}(s,t,u)=16\pi (f_0(s)+3f_1(s)\cos \theta )+\mathcal{A}_{\ell\geq 2}
\end{equation}
with the following chiral behaviour,
\begin{align}
&\re \mathcal{A}_{\ell\geq 2}\sim O(p^4),\quad \im \mathcal{A}_{\ell\geq 2}\sim O(p^8),
\\
&\re f_\ell \sim O(p^2) ,\quad \im f_\ell \sim O(p^4),
\ \ell=0,1 .
\end{align}
Having this, one can reconstruct the amplitude $\mathcal{A}(s,t,u)$ of a process $%
AB\rightarrow CD$ to $O(p^8)$ from the knowledge of the imaginary parts of $S
$ and $P$ partial waves of all the crossed amplitudes:
\begin{multline}
\mathcal{A}(s,t,u)=\mathrm{P}(s,t;u)+\Phi_0(s) \\
+\bigl[s(t-u)+(m_A^2-m_B^2)(m_C^2-m_D^2)\bigr]\Phi_1(s) \\
+\text{crossed channels}+O(p^8),
\end{multline}
where $\mathrm{P}(s,t;u)$ is a polynomial having the same $s,t,u$ symmetries
as the amplitude $\mathcal{A}(s,t;u)$ and of at most third order in
the Mandelstam variables. $\Phi _0$ and $\Phi _1$ are the
dispersive integrals of the partial waves of the $s$-channel amplitude
\begin{align}
\Phi _0(s)& =16s^3\!\!\int_{\text{thresh.}}^\infty \!\!\!\!\!dx\frac{\im f_0(x)}{%
x^3(x-s)}, \\
\Phi _1(s)& =48s^3\!\!\int_{\text{thresh.}}^\infty \!\!\!\!\!dx\frac{\im f_1(x)}{%
x^3(x-s)\lambda _{AB}^{1/2}(x)\lambda _{CD}^{1/2}(x)}\nonumber
\end{align}
and similar for the $t$- and $u$-crossed channel [$\lambda
_{AB}(x)=(s-(m_A+m_B)^2)(s-(m_A-m_B)^2)$].

The imaginary parts that enter the above expressions are obtained from the
unitarity relation,
\begin{multline}
\im f_\ell^{i\rightarrow f}(s)=\sum_k\frac{1}{S}\frac{\lambda^{1/2}_k(s)}{s} \\
\times f_\ell^{i\rightarrow k}(s)\left(f_\ell^{f\rightarrow
k}(s)\right)^{*}\theta(s-\text{thr}_k),
\end{multline}
projected on the corresponding partial waves.
The sum goes over all the possible intermediate states $k$
($S$ is a symmetry factor,  $S=2$ for undistinguishable states and $S=1$ otherwise).
In the low-energy region, and up to two-loops, these are restricted to pairs of light pseudoscalar
mesons. To the extent that we are interested in the decay region only,
we may further restrict them to intermediate $\pi\pi$ states. The contributions
from other intermediate states, like \emph{e.g.}\ $K\pi$, can be expanded in
powers of the Mandelstam variables and absorbed into the polynomial
$\mathrm{P}(s,t,u)$.

This unitarity relation and the reconstruction theorem can be used
iteratively, \emph{i.e.} starting from the LO amplitudes, we obtain the NLO results. $S$ and $P$ partial wave
projections thereof then allow to obtain the NNLO expressions.
Details of the first step will be presented
in the next section.

According to the reconstruction theorem and due to the crossing symmetry, the
two-loop representation of the $K_L\to 3\pi ^0$  amplitude looks like:
\begin{multline}
\label{amp}
\mathcal{A}_{L;00}(s,t,u)=\mathrm{P}_{L;00}+\Phi_0^{L;00}(s)+\Phi_0^{L;00}(t)\\
+\Phi_0^{L;00}(u)+O(p^8)
\end{multline}
with the polynomial
\begin{multline}
\mathrm{P}_{L;00}=C_F\bigl(A_{00}^LM_K^2+\big\{C_{00}^L[(s-s_0^L)^2] \\
+E_{00}^L[(s-s_0^L)^3]\big\} +\{s\leftrightarrow t\}+\{s\leftrightarrow u\}%
\bigr),
\end{multline}
where the centre of Dalitz plot was defined as $s_0^L=1/3M_K^2+m_0^2$ and $C_F$
corresponds to the standard normalization,
$C_F=-\frac 35V_{us}^{*}V_{ud}\frac{G_F}{\sqrt{2}}\,.$

\section{First iteration: one-loop expressions}
As already stated, we need the leading order $\pi\pi$ and
$K\pi\to\pi\pi$ scattering amplitudes. From the chiral perturbation theory we know that
at $O(p^2)$ they are represented by polynomials of at most
first order in the Mandelstam variables. Their particular choice (connected also with the particular choice of the polynomial of the reconstruction theorem)
is important since different choices can possibly lead to different convergence properties
of the chiral expansion and affect the stability of the fit to the data.
The standard choice (believed to be stable)
is the expansion in subthreshold parameters. For the $\pi\pi$ amplitude, this corresponds to
\begin{align}
\mathcal{A}_{\mathrm{LO}}^{+-;00}&= -\frac{\beta_{\pm0}}{F_\pi^2} \left(s-\frac23 m_+^2
-\frac23 m_0^2\right)-\frac{\alpha_{\pm0}m_0^2}{3F_\pi^2}\,,  \nonumber\\
\mathcal{A}_{\mathrm{LO}}^{00;00}&= \frac{\alpha_{00} m_0^2}{F_\pi^2}\,.  \label{pipi LO}
\end{align}
Another possibility would consist in choosing the scattering length and
effective range parameter as independent coefficient. It is even possible
to adjust the polynomial part $\mathrm{P}(s,t;u)$ of the $\pi\pi$ amplitude so that these coefficients retain their
physical interpretation up to two loops, just as in the non-relativistic approach \cite{bern}.
We shall study this option elsewhere \cite{plan}.
In the case of $K_L\pi^0\to\pi^0\pi^0$ and $K_L\pi^0\to\pi^+\pi^-$ amplitudes we have
\begin{align}
\mathcal{A}_{L;00}^{\mathrm{LO}} &= C_F A^L_{00} M_K^2,  \nonumber \\
\mathcal{A}_{L;+-}^{\mathrm{LO}} &= C_F \left[B^L_{+-} (s- s^L_{\pm}) + A^L_{+-} M_K^2\right]
\label{Kpi LO},
\end{align}
where $s^L_{\pm}=(M_K^2 + m_0^2 +2 m_+^2)/3$.

The $O(p^2)$ chiral perturbation theory result is reproduced by special values of the parameters ($\alpha_{00}=1$, $\beta_{\pm0}=1$, $\alpha_{\pm0}= (2 m_+^2 - m_0^2)/m_0^2$; cf.\ \cite{knecht}) and similarly for the $K\pi$ part (see \cite{bijnens}).

Using these amplitudes in the first iteration, we obtain the one-loop result for $\Phi_0^{L;00}$ of (\ref{amp}) as
\begin{equation}\label{NLO result}\begin{split}
&\Phi_0^{L;00}(s)=\frac{C_F}{2F_\pi^2} A^L_{00} M_K^2\alpha_{00} m_0^2 \bar J_0(s)\\
&-\frac{C_F}{F_\pi^2}[\beta_{\pm0}(s-\frac23 m_+^2 -\frac23 m_0^2) + \frac13
\alpha_{\pm0}m_0^2] \\
&\qquad\times[A^L_{+-}M_K^2 + B^L_{+-}(s-s^L_{+-})] \bar J_{\pm}(s)\\
&+ \mathrm{polynomial} + O(p^6).
\end{split}\end{equation}
The loop functions $\bar J$ are defined by
\begin{equation}
\bar J_{PQ}(s) = \frac{s}{16\pi^2} \int_{(m_P+m_Q)^2}^\infty dx \frac{\lambda_{PQ}^{1/2}(x)}{x(x-s)}\,,
\end{equation}
which simplifies for both masses the same to
\begin{equation}  \label{J function}
\bar J_{P}(s) = \frac{1}{16\pi^2} \left(2+\sigma_P \ln \frac{\sigma_P -1}{%
\sigma_P+1}\right).
\end{equation}

Our NLO parameterization of the amplitude (\ref{amp}) with $\Phi_0^{L;00}$ given by (\ref{NLO result}) can be now fitted to experimental data, and so we can get the $O(p^2)$ values of all the seven constants appearing in the LO parameterization of (\ref{pipi LO}) and (\ref{Kpi LO}).
It is also important to have in mind that our parameterization of the amplitude encompasses the chiral perturbation theory result \cite{bijnens} as a particular case for special values of the parameters.

As an illustration of the cusp effect in the $K_L\to3\pi^0$ decay, we plot the resulting partial decay rate for one particular choice of the subthreshold parameters in the Fig.~1.

\begin{figure}[tbp]
\begin{picture}(20,130)(0,25)
\put(0,10){\epsfig{width=0.46\textwidth,file=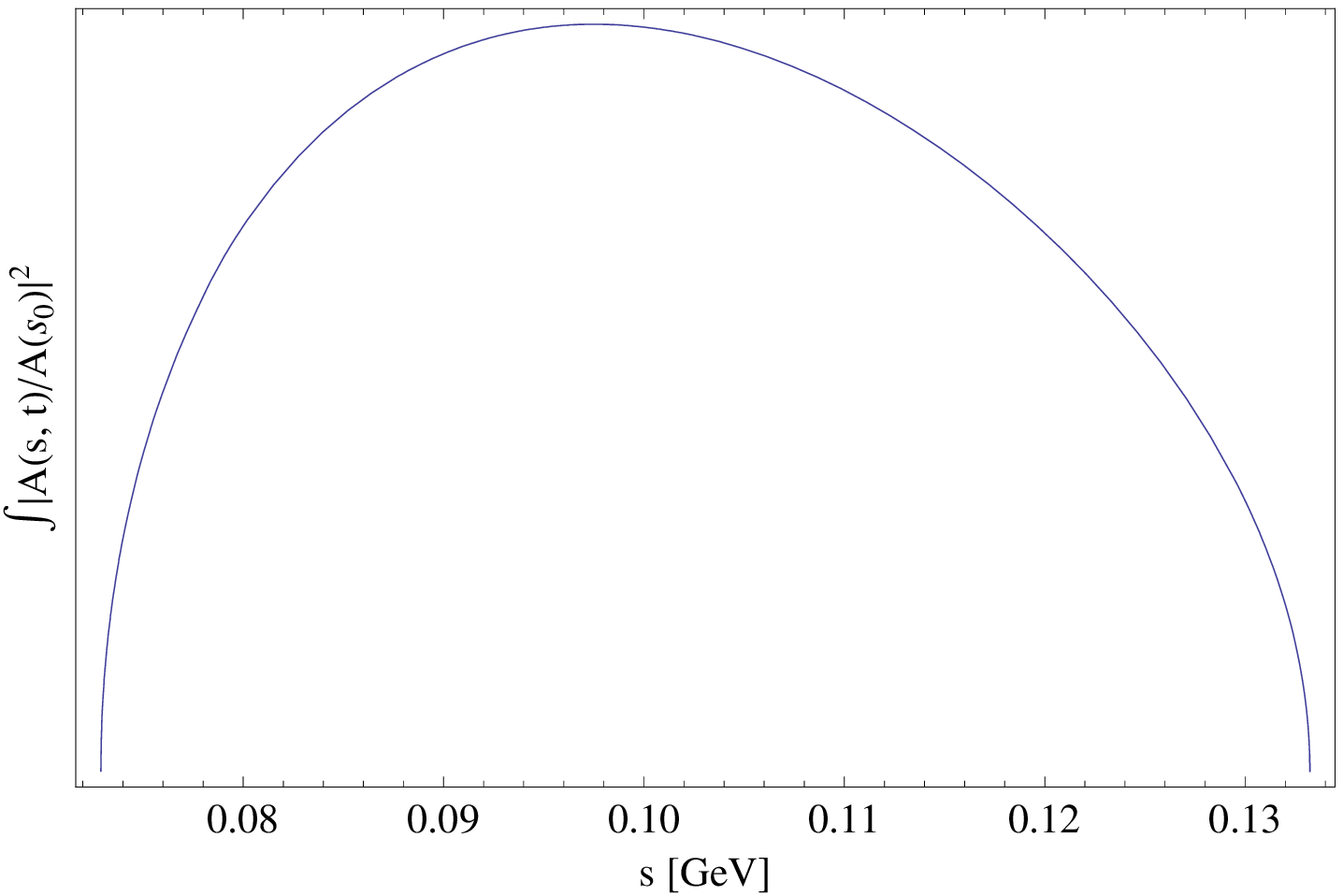}}
\put(40,32){\epsfig{width=0.25\textwidth,file=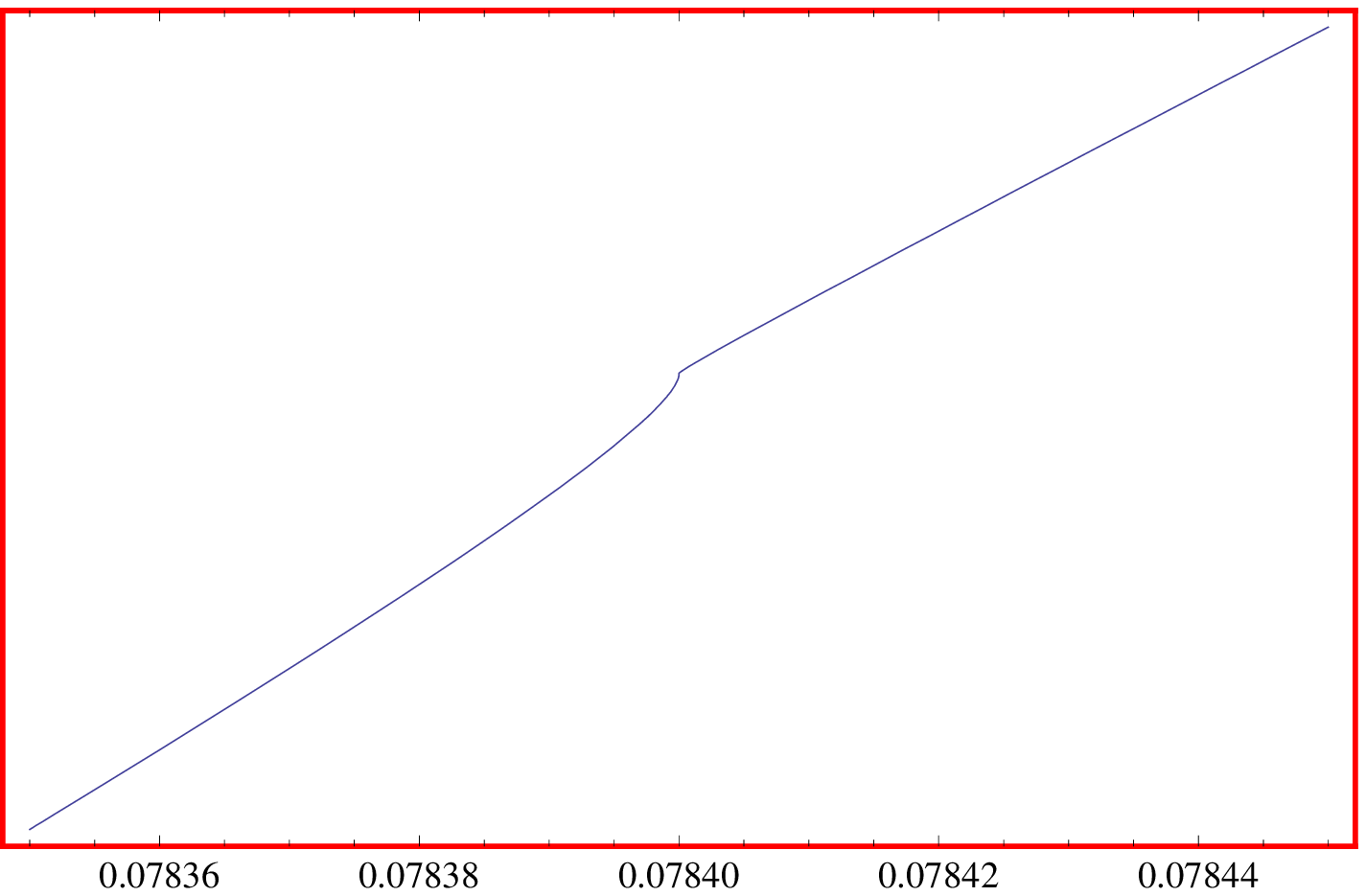}}
\put(35,110.5){\color{red}\framebox(1.2,1){}}
\end{picture}
\caption{The partial decay rate $K_L\to3\pi^0$ (in arbitrary units) as a function of the invariant mass of the $\pi^0\pi^0$ pair squared for one particular choice of parameters. Within the small frame the cusp region is zoomed.}
\end{figure}

\section{Second iteration: some remarks}

The second iteration leads to the two-loop expression of the $K\to 3\pi$
amplitude. However, a few complications have to be dealt with. As we have seen above, in order to get
the discontinuities of the function  $\Phi_0^{L;00}(x)$, we need an appropriate
analytic continuation of the unitarity relation (and therefore also an
analytic continuation of the $S$ and $P$ partial wave projections of the
amplitudes obtained by the first iteration) below the physical threshold.
In the isospin limit the solution is well known. The general method
developed in \cite{Bronzan Anisovich} is based on the careful definition of the $t$-integration contour $C(t_{+},t_{-})$
in the formula for the partial wave projections, where the endpoints are
\begin{align}
t_{\pm}(s)&=\frac 12\left(3s_0^L-s\pm \lambda_{L0}^{1/2}(s)\sigma_0\right) +
\mathrm{i}\varepsilon,  \\
\sign\varepsilon&=\sign\frac{\partial t_{\pm }(s)}{\partial
M_K^2}\,.
\end{align}
The contour $C(t_{+},t_{-})$ is defined in such a way that it avoids the intersection
with the cuts attached to branch points corresponding to the normal threshold  of the amplitude $\mathcal{A}_{L;00}(s,t,u)$ in the $t-$ and $u-$ channel. The trajectory of $t_{\pm }(s)$
and the basic types of the contour $C(t_{+},t_{-})$ in the complex $t-$plane
are depicted in \cite{Bronzan Anisovich}.  In the case of the $K_L\pi^0\rightarrow\pi^{0}\pi^{0}$ and $K_L\pi^0\rightarrow\pi^{+}\pi^{-}$
scattering, the generalization of this
prescription beyond the isospin limit is straightforward. However, this is not
the case of \emph{e.g.}\ the experimentally more interesting process
$K^{+}\pi^{-}\rightarrow\pi^0\pi^0$, where the naive application of the prescription
\cite{Bronzan Anisovich} shows some problems. Namely, for the reconstruction of the amplitude of this process beyond NLO we need to compute the partial waves of $K^{+}\pi^{-}\rightarrow\pi^+\pi^-$, where the trajectory of $t_{-}(s)$ crosses the $t-$channel cut of this amplitude instead of avoiding it, contrary to the isospin limit. In addition, the two-loop amplitude for $K^{+}\pi^{-}\rightarrow\pi^0\pi^0$ suffers from (complex) anomalous
threshold stemming from the triangle Landau singularity. The consequence is that a careful analytic
continuation of the normal dispersive integrals entering the representation of the
amplitude by means of the reconstruction theorem has to be performed. These technical
issues deserve a detailed further discussion which is however beyond the scope of this talk.

\section{Conclusion}
The cusp effect in $K\to 3\pi $ offers an interesting possibility
to extract quantitative information on $\pi\pi $ scattering lengths
from the experimental data. We have outlined a construction, in a fully relativistic
framework, of the corresponding
two-loop amplitudes, based on general properties, unitarity,
analyticity, crossing, and chiral counting for the partial waves.
Our analysis provides two parameterizations, one in terms of
the subthreshold parameters and the other, as in existing analyses,
directly in terms of scattering lengths.

\end{document}